\documentclass[runningheads]{llncs}

\usepackage{xspace}
\usepackage[svgnames,table]{xcolor}
\usepackage{graphicx}
\usepackage[allcolors=blue,colorlinks]{hyperref}
\usepackage{amsmath}
\usepackage{amsfonts}
\usepackage{wrapfig}
\usepackage{subcaption}
\usepackage{paralist}
\usepackage{underscore} %
\usepackage{cleveref}
\Crefname{section}{Sect.}{Sects.}
\Crefname{equation}{Formula}{Formulae}
\usepackage{tabularx}
\usepackage{booktabs}
\usepackage[shortcuts]{extdash}

\DeclareMathOperator{\type}{\mathsf{type}}

\title{Sound Development of Safety Supervisors}
\author{{Mario Gleirscher\inst{1}\inst{2}%
    \orcidID{0000-0002-9445-6863}%
    \thanks{Corresponding author}
    \and
    Lukas Plecher\inst{1}\orcidID{0000-0002-4991-1636}
    \and
    Jan Peleska\inst{1}\orcidID{0000-0003-3667-9775}%
    \thanks{Partially funded by the German Ministry of Economics,
      Grant Agreement~20X1908E.}}}
\institute{Mathematics \& Computer Science, University of Bremen, Bremen, Germany
  \and
  Autonomy Assurance International Programme, University of
  York, York, UK
  \email{\{gleirscher,plecher,peleska\}@uni-bremen.de}}
\titlerunning{Sound Development of Safety Supervisors}
\authorrunning{Gleirscher \and Plecher \and Peleska}

\makeatletter\begin{document}
\maketitle
\begin{abstract}
  Safety supervisors are controllers enforcing safety properties by
  keeping a system in (or returning it to) a safe state.  The
  development of such high-integrity components can benefit from a
  rigorous workflow integrating formal design and verification.
  In this paper, we present a workflow for the sound development of
  safety supervisors combining the best of two worlds, verified
  synthesis and complete testing.  Synthesis allows one to focus on
  problem specification and model validation.  Testing compensates for
  the crossing of abstraction, formalism, and tool boundaries and is a
  key element to obtain certification credit before entry into
  service.  We establish soundness of our workflow through a rigorous
  argument.
  Our approach is tool-supported, aims at modern autonomous systems,
  and is illustrated with a collaborative robotics example.

\end{abstract}

\section{Introduction}
\label{sec:intro}

\emph{Safety supervisors} (supervisors for short) are discrete-event
controllers that enforce probabilistic safety properties in modern
autonomous systems such as human-robot collaboration and autonomous
driving.  Supervisor development benefits from design and test
automation.  Because supervisors are high-integrity components, their
automation needs to be rigorously assured.  This rigor suggests
\emph{verified synthesis} for design automation and \emph{model-based
  conformance test} for test automation.  Indeed, standards and
regulations for safety-critical
systems~(e.g., \cite{ISOTS15066,ISO26262,DO178C,DO330}) require testing
as a key element to obtain certification credit before entry into
service.  To get the best of the two worlds---synthesis and test---a
crossing of the boundaries between different abstractions, formalisms,
and tools is inevitable.  This involves bridging the gap between
synthesis, the derivation of test suites from a synthesized supervisor
reference, the generation of executable code, its test and deployment
on a control system platform, and its integration into the wider
system to be put into service.

\begin{figure}[t]
  \includegraphics[width=\textwidth]{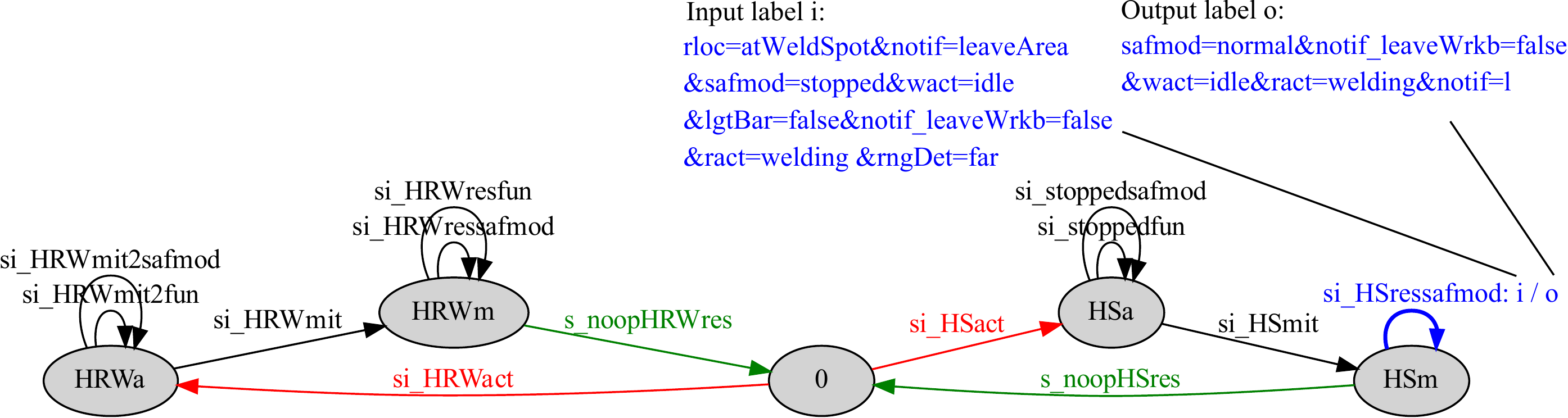}
  \caption{Example of a safety supervisor.  Nodes and edges denote
    states ($S$) and transitions ($T$) and edge
    labels signify input/output expressions ($(i,o)\in\Sigma$).}
  \label{fig:testref}
\end{figure}

To resolve this challenge, we propose a \emph{rigorous workflow for
  the sound development, in particular, the synthesis and complete
  test, of safety supervisors}.
\label{sec:runn-example-robot}%
Prior to explaining our workflow, we illustrate safety supervision by
example of an operator collaborating with a robot on a welding and
assembly task in a workcell with a spot
welder~\cite{Gleirscher2020-SafetyControllerSynthesis,%
  Gleirscher2021-VerifiedSynthesisSafety}.  These actors perform
dangerous actions~(e.g., robot movements, welding steps) possibly
reaching hazardous states~(e.g., operator near the active spot welder,
$\mathit{HS}$; operator and robot on the workbench, $\mathit{HRW}$).
To reduce accident likelihood, such states~(\Cref{fig:testref}) need
to be reacted to.  These reactions are under the responsibility of a
supervisor (\Cref{fig:testref}) enforcing probabilistic safety
properties of the workcell, such as ``accident $a$ is less likely than
probability $\mathit{pr}_a$'' or ``hazard $h$ occurs less likely than
$\mathit{pr}_h$''.  The supervisor's behavior comprises
\begin{inparaenum}[(i)]
\item the detection of \textcolor{red}{\emph{critical events}} (e.g.,
  \texttt{si\_HSact}),
\item the performance of \emph{mitigation actions} (e.g.,
  \texttt{si\_stoppedfun}) to react to such events and reach a safe
  state (e.g., $\mathit{HSm}$), and
\item avoiding a paused task or degraded task performance, the
  execution of \textcolor{Green}{\emph{resumption actions}} (e.g.,
  \textcolor{blue}{\texttt{si_HSressafmod}}) to resolve the event and
  to return to a safe but productive state~(here, $0$).
\end{inparaenum}

Extending our previous work
\cite{Gleirscher2021-CompleteTestingSafety}, we propose the derivation
of symbolic finite state machines (SFSM) used as \emph{test
  references} for model-based testing with complete methods.  The
resulting test suites allow for a \emph{proof} of conformance (i.e.,
observational equivalence) between the generated supervisor code and
the reference.  The hypotheses to be fulfilled to guarantee test suite
completeness can be checked by simple static analyses of the
supervisor code.  We provide tool support for both these steps,
explain the tool qualification obligations, and present a rigorous
argument by applying Hoare logic on the workflow
to obtain certification credit for the supervisor, on the basis of the
development, verification, and validation process, and the artifacts
produced in each workflow stage.

 \Cref{sec:approach} summarizes the proposed workflow.
\Cref{sec:workflow} details each workflow stage.  In
\Cref{sec:workflow-qualif-concept}, we argue for the soundness of our
approach, make our assumptions explicit, and show how we reduce
several error possibilities in our workflow.  \Cref{sec:related-work}
summarizes related work.  We add concluding remarks in
\Cref{sec:conclusion}.

\section{Overview of the Workflow}
\label{sec:approach}

\begin{figure}
\includegraphics[width=\linewidth]{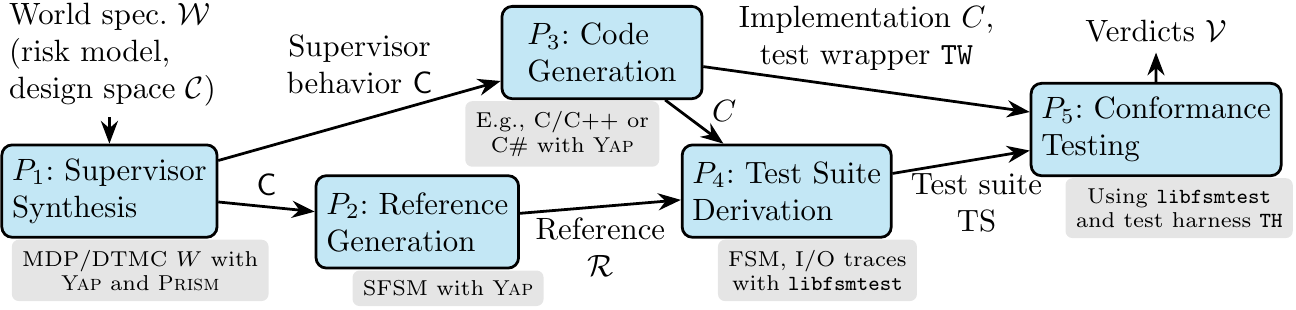}
  
  \caption{Stages $P_i$ and artifacts of the proposed supervisor development workflow}
  \label{fig:workflow}
  
\end{figure}

\Cref{fig:workflow} shows our workflow.  In $P_1$, we construct a
\emph{stochastic world model}~$W$ describing the behaviors of all
relevant actors~(e.g., humans, robots, equipment) and the supervisor.
$W$ includes a range of controller behaviors some of which---when
implemented correctly---guarantee that the risk in the actual world is
acceptable, provided that the stochastic assumptions have been made to
the safe side.  This range is denoted as the supervisor \emph{design
  space} $\mathcal{C}$.  $P_1$ adopts our work
\cite{Gleirscher2021-VerifiedSynthesisSafety,%
  Gleirscher2020-SafetyControllerSynthesis} based on policy synthesis
for Markov decision processes~(MDPs)
\cite{Kwiatkowska2011-PRISM4Verification,%
  Kwiatkowska2007-StochasticModelChecking} and selects controller
behaviors from $\mathcal{C}$ that meet requirements $\phi$~(see example in
\Cref{sec:runn-example-robot}) specified in probabilistic computation
tree logic (PCTL; \cite{Kwiatkowska2011-PRISM4Verification}) and
verified of~$W$ (and $\mathcal{C}$).  While maintaining the constraint
$\phi$, the synthesis procedure applies objectives (e.g., maximum
performance, minimum cost) and selects an optimal \emph{supervisor
  behavior} $\mathsf{C}\in\mathcal{C}$.

In $P_2$, $\mathsf{C}$ is transformed into a \emph{test reference}
$\mathcal{R}$, an SFSM~\cite{DBLP:conf/icst/Petrenko16} whose control
states are called risk states and transitions are labeled with
input/output~(I/O) pairs (e.g., \Cref{fig:testref}).  The input
alphabet of~$\mathcal{R}$ specifies the events (i.e., state changes
in~$W$) observed and the output alphabet the signals issued to
$W$ by the supervisor.  Input labels model guard conditions that,
when holding \texttt{true} of a world state, enable or trigger their
transition~(e.g., \texttt{si_HSressafmod} in \Cref{fig:testref}).

In $P_3$, $\mathsf{C}$ is translated into a software component $C$
executed by the control system of a robotic or autonomous system.
Following an embedded systems tradition, our example uses C++ as the
target language for $C$.  Moreover, we assume that SFSMs have a
simpler semantics than the executable code.

In $P_4$, a complete test suite $\mathrm{TS}$ for checking $C$
against $\mathcal{R}$ is generated, using a general input equivalence
class testing strategy for SFSMs~\cite{peleskasttt2014}. The term
\emph{complete} means that---provided certain hypotheses
hold---$\mathrm{TS}$ will (i) accept every $C$ whose behavior
is represented by an SFSM which is observationally equivalent to
$\mathcal{R}$, and (ii) reject every non-equivalent implementation.  It
is shown below that these hypotheses are fulfilled by $C$, so
that \emph{passing} $\mathrm{TS}$ corresponds to a \emph{proof} of
\emph{observational equivalence}.

In $P_5$, $\mathrm{TS}$ is run against $C$ to record inputs,
associated outputs, and the \emph{verdicts} $\mathcal{V}$ obtained for
each test case in $\mathrm{TS}$.  A generated test harness
$\texttt{TH}$, emulating the target platform, acts as the \emph{test
  oracle} by comparing the I/O traces observed by the 
wrapper $\texttt{TW}$ during execution of
$\mathrm{TS}$ to the traces expected according to~$\mathcal{R}$.

\section{Workflow Stages}
\label{sec:workflow}

This section details the five workflow stages outlined in
\Cref{sec:approach} and \Cref{fig:workflow}.

\subsection{$P_1$: Risk-informed Supervisor Synthesis}
\label{sec:deriv-test-refer}

This section summarizes \cite{Gleirscher2021-VerifiedSynthesisSafety}
for the construction of $W$ and selection of the supervisor
behavior~$\mathsf{C}$ from $W$.
Below, $\cal P;Q$, $\cal P\sqcap Q$, $\cal P^{\star}$, and
$\cal P\setminus Q$ signify the sequential composition of the
behaviors $\cal P$ and $\cal Q$, non-deterministic choice between
$\cal P$ and $\cal Q$, non-deterministic but finite repetition of
$\cal P$, and the removal of $\cal Q$ from $\cal P$,
respectively.  Moreover, $\cal P\sqsubseteq Q$ denotes that
$\cal Q$ refines $\cal P$ while also preserving progress, and
$\cal P=Q$ that $\cal P$ and $\cal Q$ are observationally equivalent.

First, we specify $W$ as a probabilistic program
$\mathcal{W} = (\mathcal{E};\mathcal{C})^{\star}$ alternating between an
\emph{environment}~$\mathcal{E}$ (e.g., robot, operator) and a
\emph{supervisor design space} $\mathcal{C}$.  Given the set
$A_{\mathcal{W}} = A_{\mathcal{E}} \cup A_{\mathcal{C}}$ of all
probabilistic commands of~$\mathcal{W}$ (e.g., operator or robot moves,
welding steps, \Cref{sec:runn-example-robot}), we require $\mathcal{E}$ and
$\mathcal{C}$ to be refinements of the command bundles
$(\sqcap_{\alpha\in A_{\mathcal{E}}} \alpha)\sqsubseteq\mathcal{E}$ and
$(\sqcap_{\alpha\in A_{\mathcal{C}}} \alpha)\sqsubseteq\mathcal{C}$.  With
$\mathit{idle}\in A_{\mathcal{C}}$, the world without the supervisor
can be generated by
$\mathcal{W}\setminus\mathcal{C} = (\mathcal{E};\mathit{idle})^{\star}$.

Let a set $V$ of finite-sorted variables, $\type v$ the
sort of $v\in V$, and the universe
$\mathbb{U} = \bigcup_{v\in V}\type v$.  A state is a
valuation function $s\colon V\to\mathbb{U}$ with
$\forall v\in V\colon s(v)\in\type v$.  Map restriction
$\cdot|_{V'}$ restricts a (set of) state(s) from
$V\to\mathbb{U}$ to $(V\cap V')\to\mathbb{U}$.
Let $\mathbb{S}$ be the set of all states.
Given $\alpha\colon\mathbb{S}\to 2^{\mathbb{S}}$ for
$\alpha\in A_{\mathcal{W}}$, by inductively applying $\mathcal{W}$ to an
initial state $s_0\in\mathbb{S}$, we obtain an MDP $W$
and the set $\mathcal{S}\subseteq\mathbb{S}$ of states reachable by
$\mathcal{W}(s_0)$.\footnote{Below, we mostly abbreviate
  ${\cal P}(s_0)$ to $\cal P$ when referring to the
  transition relation of a program $\cal P$ executed from initial state
  $s_0\in\mathbb{S}$. So, $\cal P\models\phi$ actually means
  ${\cal P}(s_0)\models\phi$.}  $\mathcal{S}$ is labeled
with atomic propositions.
$W$ is a labeled transition system whose transition relation
encodes non-deterministic and probabilistic choice.  MDPs can model
\emph{uncertainties} about uncontrolled or non-modeled aspects in
$\mathcal{E}$.  See, e.g., \cite{Kwiatkowska2011-PRISM4Verification} for
further details.

\begin{figure}[t]
  \begin{subfigure}{.48\linewidth}
\includegraphics[width=.8\linewidth]{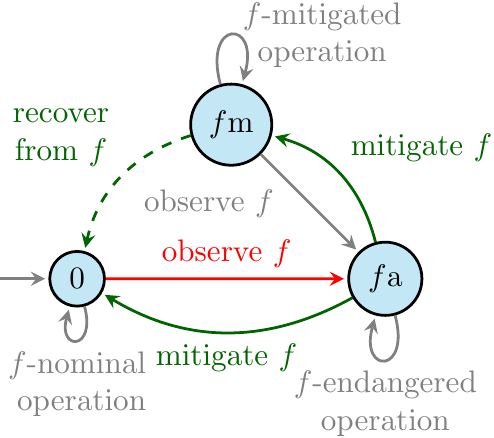}
    \caption{Phases and transitions of factor $f$}
    \label{fig:factor}
  \end{subfigure}
  \hspace{-1em}
  \begin{subfigure}{.48\linewidth}
\includegraphics[width=\linewidth]{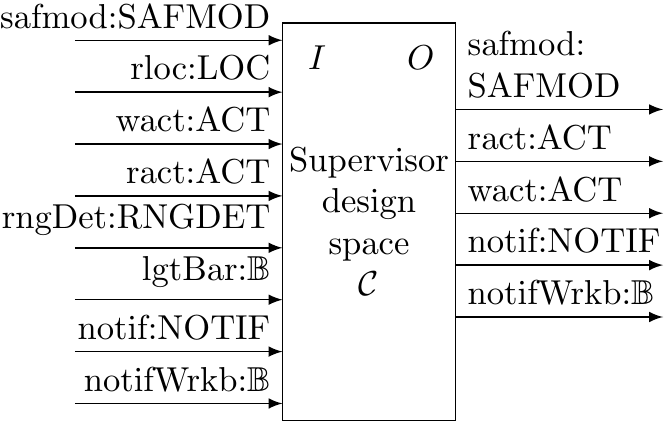}
    \caption{Syntactic interface between $\mathcal{E}$ and $\mathcal{C}$}
    \label{fig:ctr-interface}
  \end{subfigure}
  
  \caption{Risk factor and supervisor interface}
  \label{fig:factor-interface}
  \vspace{-1em}
\end{figure}

We model risk using a set $F\subset V$ of
$\mathbb{P}$-sorted variables describing the critical events in~$W$
as \emph{risk factors}~\cite{Gleirscher2021-RiskStructuresDesign}.  A
risk factor $f\in F$ (\Cref{fig:factor}) has at least three
phases $\mathbb{P}=\{0,a,m\}$ as well as phase transitions in
$A_{\mathcal{C}}$ modeling the life-cycle of handling $f$, for
example, from \emph{inactive} ($0$) to \emph{active} ($f$a) to
\emph{mitigated} ($f$m) and back to inactive.  The example in
\Cref{sec:runn-example-robot} uses three factors,
$F = \{\mathit{HS}, \mathit{HC}, \mathit{HRW}\}$.  That way,
$F$ induces a notion of \emph{risk state} in $\mathcal{S}$.
$\mathcal{C}$ can then be associated with states
$S=\mathcal{S}|_{F}\subseteq (F\to\mathbb{P})$,
which represent the possible control states of any supervisor policy
$\mathsf{C}$, test reference $\mathcal{R}$, and implementation $C$.

Freedom of choice in $\mathcal{E}$ and $\mathcal{C}$ is resolved by refining
indeterminacy in~$\mathcal{E}$ through uniformly distributed probabilistic
choice and in $\mathcal{C}$ by deriving a \emph{policy}, a choice
resolution for each state in $\mathcal{S}$ where the supervisor is
enabled and can make a decision.  The result is a discrete-time Markov
chain~(DTMC) $ (\mathsf{E}; \mathsf{C})^{\star}\sqsupseteq\mathcal{W}$, a labeled
transition system \emph{without indeterminacy}.  Here, policy
derivation involves both sub-setting~$\mathcal{C}$ using PCTL constraints
$\phi$ and Pareto-optimizing multiple
objectives~\cite{Kwiatkowska2011-PRISM4Verification}.  Any resulting
optimal DTMC $(\mathsf{E}; \mathsf{C})^{\star}$ is thus verified against
$\phi$, establishing $(\mathsf{E}; \mathsf{C})^{\star}\models\phi$, and
includes the selected behavior $\mathsf{C}\sqsupseteq\mathcal{C}$.

\subsection{$P_2$: Test Reference Generation}
\label{sec:test-ref-gen}

For the translation of $\mathsf{C}$ into a \emph{test reference}
$\mathcal{R}$, we define the I/O alphabet
$\Sigma\subseteq\Sigma_{I}\times\Sigma_{O}$ with
$\Sigma_{I} = I\to\mathbb{U}$ and
$\Sigma_{O} = O\to\mathbb{U}$ for monitored and
controlled variables $I,O\subseteq V$, resulting
in the \emph{syntactic
  interface}~\cite{Broy2010-LogicalBasisComponent} of $\mathcal{C}$.  This
interface (\Cref{fig:ctr-interface}) defines the nature of the changes
in $\mathcal{E}$ that any $\mathsf{C}$ can observe and perform.  We require
$(I\cup O)\cap F=\emptyset$ but allow an overlap of
$I$ and $O$.

To obtain $\mathcal{R}$ as an operation refinement %
of the parallel composition of all factors in $F$, we translate
the $\mathsf{C}$-fragment of the transition relation of the DTMC
$(\mathsf{E};\mathsf{C})^{\star}$ into a deterministic SFSM
$\mathcal{R} = (S,\Sigma,T,\overline{s})$ with states
$S=\mathcal{S}|_{F}$, the transition relation
$T\subseteq S\times\Sigma\times S$, and the
initial state $\overline{s}\in S$ congruent
with~$s_0$~(i.e.,
$\forall f\in F\colon s_0(f)=\overline{s}(f)$), usually
$\overline{s}=0$.
Given a DTMC transition $(s,s')\in\mathcal{S}\times\mathcal{S}$, the
source state $s$ is mapped to input
$i=s|_{I}\in\Sigma_{I}$~(the observed event) and risk
state $r\in S$.  The control and state updates
$o=s|_{O}\in\Sigma_{O}$ and $r'\in S$ are
derived from the difference in the controlled and factor variables
$O\cup F$ between $s$ and the target state~$s'$.
\Cref{fig:testref} shows an example of $\mathcal{R}$.
Finally, $\mathcal{R}$ is provided in a format readable by
\texttt{libfsmtest}\xspace\footnote{Licensed according to MIT license
  \url{https://opensource.org/licenses/MIT}. Source code available
  under
  \url{https://bitbucket.org/JanPeleska/libfsmtest}.}~\cite{libfsmtest}
for test suite generation (\Cref{sec:test-strategy}).

\vspace{-1em}

\subsection{$P_3$: Code Generation}
\label{sec:code-gen}

\begin{figure}[t]
  \begin{subfigure}{.8\linewidth}
\includegraphics[width=\linewidth]{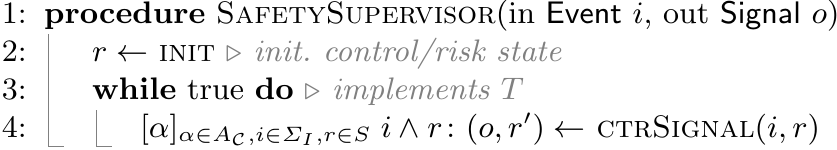}
    \caption{Pseudo code of a generic supervisor} 
    \label{alg:codepattern}
  \end{subfigure}
  
  \begin{subfigure}{\linewidth}
\includegraphics[width=\linewidth]{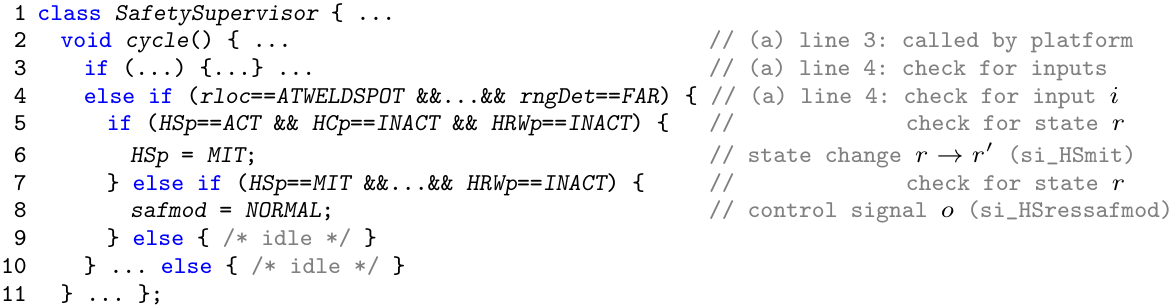}
    
    \caption{Fragment of the supervisor implementation
      $C_{C++}$ for \Cref{fig:testref}}
    \label{lst:ctrimpl}
  \end{subfigure}
  
  \caption{Supervisor pseudo code and implementation fragment}
  
\end{figure}

Independent of $P_2$, the $\mathsf{C}$-fragment of the transition
relation of $(\mathsf{E};\mathsf{C})^{\star}$ is translated to an
implementation $C$ (\Cref{alg:codepattern}).  Similar to the
translation to $\mathcal{R}$ (\Cref{sec:test-ref-gen}), every transition
of $\mathsf{C}$ is mapped\footnote{Note that the elements of
  $\Sigma_{I}$ and $S$ are used as propositions in
  guard conditions.}  to a guarded command
$[\alpha]\,i\land r\colon (o,r')\leftarrow\textsc{ctrSignal}(i,r)$
with $r,r'\in S, (i,o)\in\Sigma$, and action name
$\alpha\in A_{\mathcal{C}}$ derived from $F$, $r$, and~$r'$.
$C$ is intentionally simple (e.g., flat branching structure)
and wrapped into platform-specific code (not shown) for data
processing and communication.  \Cref{lst:ctrimpl} depicts a fragment
of $C$ for our running example.

As opposed to $\mathcal{R}$, the representation of $C$ can vary
significantly.  For instance, in \Cref{lst:ctrimpl}, we consider a C++
component for a low-level real-time implementation.  Alternatively,
one may want to derive VHDL or Verilog HDL to synthesize an
FPGA.\footnote{VHSIC or Verilog hardware description language (VHDL or
  Verilog HDL); field-programmable gate array (FPGA)} In
\cite{Gleirscher2021-VerifiedSynthesisSafety}, we consider a C\#
component used in a simulation of $W$ in a Robot Operating
System-enabled digital twinning
environment~\cite{Douthwaite2021-ModularDigitalTwinning}.  $\mathcal{R}$
can thus be shared between a varying $C_{\mathrm{C++}}$,
$C_{\mathrm{C\#}}$, and $C_{\mathrm{VHDL}}$.  The only
difference on the testing side is in the I/O wrapper used in the test
harness~(\Cref{sec:test-strategy}) to deliver the inputs to $C$
and record the outputs of $C$.
The translations into $\mathcal{R}$ and $C$ and the generation of
the test wrapper are performed with the \textsc{Yap}\xspace tool.\footnote{Features
  and examples
  available in \textsc{Yap}\xspace 0.8+, \url{https://yap.gleirscher.at}.}

\subsection{$P_4$: Test Suite Generation}
\label{sec:test-strategy}

We avoid the costly verification of the (potentially changing) code
generator used in $P_3$.  Instead, we generate a conformance test
suite that, when passed, corresponds to a \emph{correctness proof} of
$C$.  The SFSM reference model~$\mathcal{R}$
(\Cref{sec:test-ref-gen}) and the supervisor implementation $C$
(\Cref{sec:code-gen}) allow us to apply a model-based conformance
testing approach for verifying $\mathcal{R}=C$, the observational
equivalence of $\mathcal{R}$ and $C$.  Indeed,
\emph{complete} test methods enable us to prove that the system under
test~(SUT) conforms to a given reference model under certain
hypotheses~\cite{PeleskaHuangLectureNotesMBT}.  A complete test suite
is derived from $\mathcal{R}$ for $C$ as follows.

\emph{Step~1.} Since $\mathcal{R}$ is a deterministic SFSM which outputs
a finite range of control signals, we can apply the equivalence class
testing theory of Huang et al.~\cite{peleskasttt2014}, which has been
elaborated for a more general class of Kripke structures including
SFSMs like $\mathcal{R}$. Moreover, the guard conditions $c$ in
$\mathcal{R}$ are even mutually exclusive (see the construction of $T$ in
\Cref{sec:test-ref-gen} and \Cref{sec:arg-testderiv}).
Therefore, an \emph{input equivalence class} corresponds to the set of
input valuations $s|_{I}\colon I\to\mathbb{U}$ satisfying
a specific guard condition $c$.  We write $s\models c$ if the guard
condition $c$ evaluates to true, after having replaced all occurrences
of input variables $v\in I$ in $c$ by value $s(v)$.  In any SFSM
state, all members of an input class produce the same output
values. We use the guard formula $c$ itself as an identifier of the
associated input equivalence class
$\{ s|_{I}\colon I\to\mathbb{U}~|~s\in\mathbb{S} \land s\models c
\}$.

\emph{Step~2.} With identifiers $c$ as finite input alphabet and the
finite set of possible outputs as output alphabet, ${\mathcal{R}}$ is
abstracted to a minimal, deterministic, finite state machine (FSM)
$\mathfrak{A}({\mathcal{R}})$ (\Cref{fig:commdia} left) with I/O alphabet
$\mathfrak{A}({\mathcal{R}})_{I}\times\mathfrak{A}({\mathcal{R}})_{O}$.  Using
$\mathfrak{A}({\mathcal{R}})$ as a reference model, the
H-Method~\cite{DBLP:conf/forte/DorofeevaEY05} is applied to derive a
complete FSM test suite $\mathrm{TS}^{\mathit{fsm}}$. The completeness of
$\mathrm{TS}^{\mathit{fsm}}$ is guaranteed provided that the true behavior of
$C$ can be abstracted to a deterministic FSM with at most
$m\ge n$ states, where $n$ is the number of states in
$\mathfrak{A}({\mathcal{R}})$.
The generation of $\mathrm{TS}^{\mathit{fsm}}$ is performed by means of the open
source library \texttt{libfsmtest}\xspace, which contains algorithms for model-based
testing against FSMs~\cite{libfsmtest}.

\emph{Step~3.} Test suite $\mathrm{TS}^{\mathit{fsm}}$ is translated to an SFSM
test suite $\mathrm{TS}$ by selecting an input valuation
$s|_{I}$ %
from each input equivalence class $c$ and replacing each test case
$t = c_1\dots c_k\in\mathrm{TS}^{\mathit{fsm}}$ (i.e.,~sequence of input class
identifiers) by the sequences $s_1|_{I}\dots s_k|_{I}$ of
valuation functions.  It has been shown that $\mathrm{TS}$ is complete
whenever $\mathrm{TS}^{\mathit{fsm}}$ is~\cite{peleskasttt2014}.

\subsection{$P_5$: Conformance Testing}
\label{sec:conftesting}

\begin{figure}[t]
  \centering
\includegraphics[width=\linewidth]{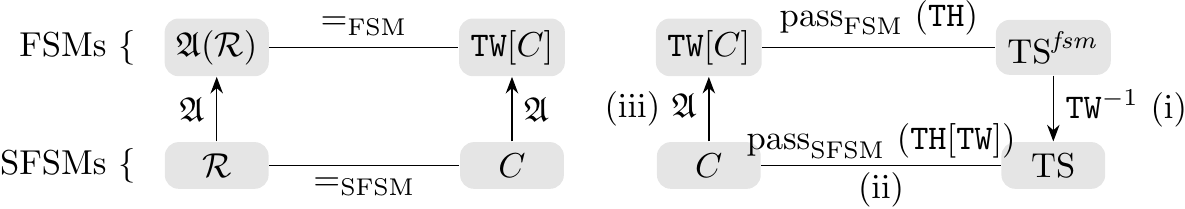}
  \caption{Commuting diagrams reflecting the observational equivalences
    ($=_{\mathrm{FSM}}$, $=_{\mathrm{SFSM}}$), pass relations
    ($\mathrm{pass}_{\mathrm{FSM}}$, $\mathrm{pass}_{\mathrm{SFSM}}$), and
    abstraction map
    ($\mathfrak{A} = \texttt{TW}$)}
  \label{fig:commdia}
\end{figure}

\texttt{libfsmtest}\xspace provides a generic \emph{test harness} $\texttt{TH}$ for
executing test suites generated from FSMs against SUTs.  For verifying
$C$, $\texttt{TH}$ executes the complete FSM test suite
$\mathrm{TS}^{\mathit{fsm}}$. To this end, the harness uses a \emph{test wrapper}
$\texttt{TW}$ for (i)~refining each
$c\in\mathfrak{A}({\mathcal{R}})_{I}$ in a test case
$t\in\mathrm{TS}^{\mathit{fsm}}$ to a concrete input valuation
$s|_{I}\in\Sigma_{I}$ of $C$, (ii)~calling
$C$ to perform one control step, and (iii)~abstracting
$C$-output valuations $s|_{O}\in\Sigma_{O}$
back to $y\in\mathfrak{A}({\mathcal{R}})_{O}$ (\Cref{fig:commdia}
right).  Test harness, wrapper, and supervisor code $C$ are
compiled and linked; this results in a test executable
$\texttt{TH}[\texttt{TW}[C]]$. The component
$\texttt{TW}[C]$ of the test executable acts like an FSM over
alphabets $\mathfrak{A}({\mathcal{R}})_{I}$,
$\mathfrak{A}({\mathcal{R}})_{O}$.  This FSM interface is used by the
test harness to stimulate $\texttt{TW}[C]$ with inputs from
$\mathfrak{A}({\mathcal{R}})_{I}$ and to check whether the outputs
$y\in\mathfrak{A}({\mathcal{R}})_{O}$ conform to the outputs expected
according to reference FSM $\mathfrak{A}({\mathcal{R}})$.

$\texttt{TW}$ implements two bijections,
$\gamma\colon\mathfrak{A}({\mathcal{R}})_{I} \to \Sigma_{I}$
for step (i) and
$\omega\colon\Sigma_{O} \to \mathfrak{A}({\mathcal{R}})_{O}$ for step (iii).
Map $\gamma$ satisfies
$\forall c\in\mathfrak{A}({\mathcal{R}})_{I}: \gamma(c)\models c$, and
$\omega$ fulfills
$\forall s\in\mathbb{S}\colon s|_{O}\in C_{O}\Rightarrow s\models
\omega(s|_{O})$. These mappings ensure that the diagrams in
\Cref{fig:commdia} are both commutative, that is, the execution of
$\mathrm{TS}^{\mathit{fsm}}$ against $\texttt{TW}[C]$ by $\texttt{TH}$
results in the execution of $\mathrm{TS}$ against $C$ by
$\texttt{TH}[\texttt{TW}[\cdot]]$.

Consider, for example, transition
\textcolor{blue}{\texttt{si_HSressafmod}} in
\Cref{fig:testref}. Function $\gamma$ maps $\mathfrak{A}({\mathcal{R}})$-input
\textcolor{blue}{\texttt{rloc=atWeldSpot\&\dots\&rngDet=far}} to $C$-input valuation
$\{ {\tt rloc} \mapsto {\tt atWeldSpot},\dots, {\tt rngDet} \mapsto {\tt far} \}$. The wrapper implements this simply by assignments {\tt rloc=atWeldSpot;\dots;rngDet=far;}.
If the corresponding $C$-step produces any output valuation in 
$\{ {\tt safmod} \mapsto y_1, \dots, {\tt notif} \mapsto y_k   \}$, this is mapped by $\omega$
to $\mathfrak{A}({\mathcal{R}})$-output 
\textcolor{blue}{\texttt{safmod=$y_1$\&\dots\&notif=$y_k$}}.

\section{Workflow Soundness} 
\label{sec:workflow-qualif-concept}

It remains to be shown that the stages $P_1$ to $P_5$ establish the
chain of refinements
\begin{equation}
  \label{eq:ref-chain}
  \mathcal{W} \stackrel{P_1}{=}
  (\mathcal{E};\mathcal{C})^{\star} \stackrel{P_1}{\sqsubseteq}
  (\mathcal{E};\mathsf{C})^{\star} \stackrel{P_1}{\sqsubseteq}
  (\mathsf{E};\mathsf{C})^{\star} \stackrel{P_2}{=}
  (\mathsf{E};\mathcal{R})^{\star} \stackrel{P_3,P_4,P_5}{=}
  (\mathsf{E};C)^{\star}\;.
\end{equation}
Through model checking of $(\mathsf{E};\mathsf{C})^{\star}\models\phi'$, our
notion of $\sqsubseteq$ preserves \emph{trustworthiness} (i.e.,
$\phi$, \Cref{sec:deriv-test-refer}) comprising
world \emph{safety} and supervision \emph{progress}.
After having established through $P_1$ that $\mathsf{C}$ is trustworthy,
trustworthiness of $\mathcal{R}$ and $C$ can be inferred from the
two observational equivalences in \eqref{eq:ref-chain}.  Furthermore,
$C$ remains trustworthy as long as
$(\mathcal{E};C)^{\star}(s)\subseteq\mathcal{S}$, implying that
$\mathcal{E}$ and $C$ are re-initialized congruently in
$s\in\mathcal{S}$.  So, the main objective of proving workflow
soundness is to establish that \eqref{eq:ref-chain} holds.

\begin{figure}[t]
  \centering
  \includegraphics[width=\textwidth]{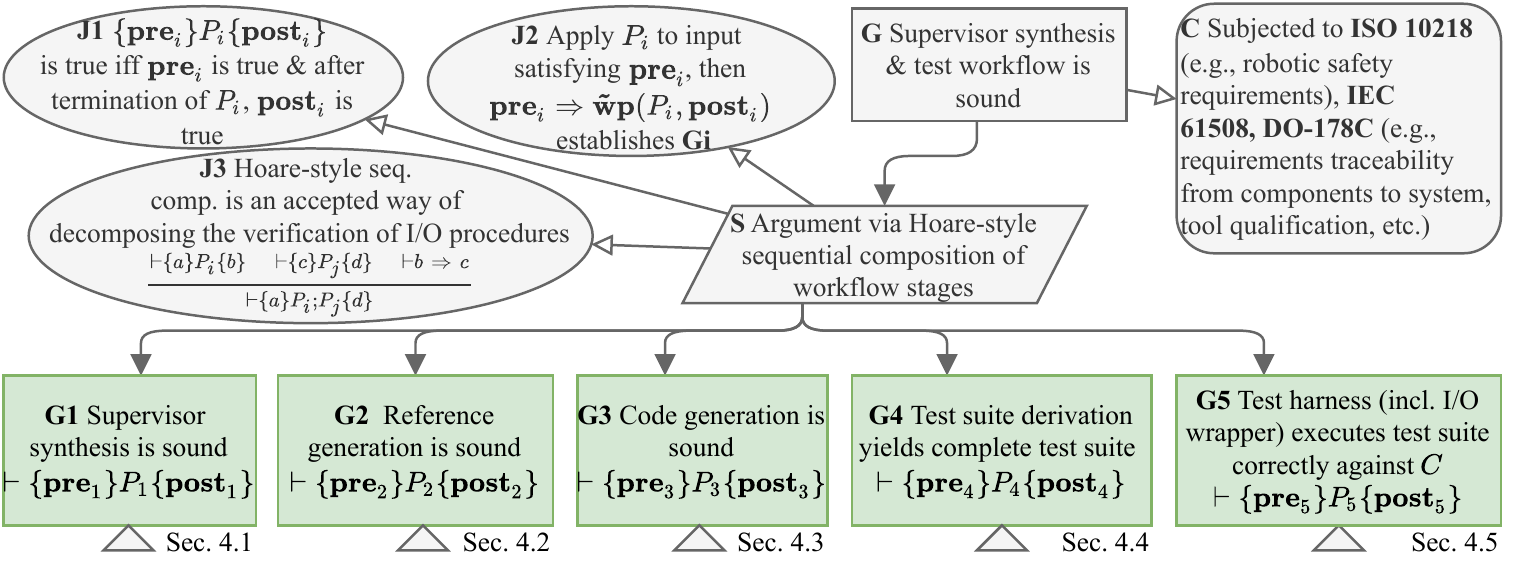}
  \caption{Overview of the workflow assurance case in goal structuring
    notation~\cite{GSN2011}}
  \label{fig:workflow-gsn}
\end{figure}

Following %
safety-critical control standards (e.g., DO-178C/330
\cite{DO178C,DO330}, IEC 61508 \cite{iec61508},
ISO~10218~\cite{ISO10218}), we argue for the soundness of our workflow
(\textbf{G}) and the refinement chain~\eqref{eq:ref-chain}.  Our
argument for \textbf{G} (top-level in \Cref{fig:workflow-gsn})
aims at ruling out that a faulty $\mathcal{R}$, test theory, $\mathrm{TS}$
generator, or test harness mask errors inside $C$.  Arguing for
\textbf{G} is known as \emph{verification of the verification
  results}.  By semi-formal weakest precondition ($\mathrm{\mathbf{\tilde{w}p}}$)
reasoning~(\textbf{J2}), we establish a Hoare triple~(\textbf{J1}) for
each workflow stage (\textbf{G1} to \textbf{G5}) and implication
relationships (\textbf{J3}) between the post- and
preconditions~(\Cref{tab:workflow-ag}) of these triples to
establish~\textbf{G}, the soundness of the sequential composition
$P_1;P_2;P_3;P_4;P_5$ (\Cref{fig:workflow}).

\begin{table}
  \caption{Overview of pre- and post-conditions of the workflow stages}
  \label{tab:workflow-ag}
  \aboverulesep=0.2mm
  \belowrulesep=0.2mm
  \tabcolsep=.5mm
  \begin{tabularx}{\textwidth}{cXX}
    \toprule
    \rowcolor{gray!20}
    $P_i$ & $\mathbf{pre}_{i}$ & $\mathbf{post}_{i}$
    \\\midrule
    1
    & Requirements $\phi$ are complete and world model $W$ is trustworthy.
    & Chosen supervisor behavior $\mathsf{C}$ can be trusted.
    \\\midrule
    2
    & Supervisor behavior $\mathsf{C}$ is trustworthy, deterministic, factor-complete,
    and $s_0$(initial state)-compatible.
    & Reference $\mathcal{R}$ is deterministic, 
    input-complete up to idle self-loops, $s_0$-compatible, and accepted by $P_4$.
    \\\midrule
    3
    & $\mathbf{post}_{1}$ and syntactic interface is defined.
    & Generated code $C$ is compatible with the test harness and
    statically analyzable.
    \\\midrule
    4
    & The complete testing theory is correct,
    the prerequisites
    for applying the selected test generation method are fulfilled,
    and the test suite validator $\text{Val}_H$ is correct.
    & The generated FSM test suite $\mathrm{TS}^{\mathit{fsm}}$ is complete for checking observational
    equivalence between   $\mathfrak{A}(\mathcal{R})$ and FSMs over the same alphabet with at most
    $m$ states, or $\text{Val}_H$ will indicate an error.
    \\\midrule
    5
    & $\mathbf{post}_{4}$, $\texttt{TW}$ is correct, and $\texttt{TH}$ is correct.
    & $C$ passes test suite if and only if it is observationally
    equivalent to $\mathcal{R}$.
    \\\bottomrule
  \end{tabularx}
\end{table}

\subsection{Assurance of $P_1$: Supervisor Synthesis}
\label{sec:arg-synthesis}

For $P_1$, we need to establish
\begin{equation}
  \label{eq:ht-synthesis}
  \{\mathbf{pre}_{1}\} P_1 \{\mathbf{post}_{1}\} \;.
  \tag{G1}
\end{equation}
$\mathbf{pre}_{1}$: For the synthesis stage to yield a trustworthy result, we
first identify a complete list $\phi$ of well-formedness properties
and supervisor factor handling requirements ($\mathbf{pre}_{1.1}$) for the MDP
$W$.  Completeness of $\phi$ relies on the completeness of
$F$, the latter on state-of-the-art hazard analysis and risk
assessment.  We are further provided that $W$ is a trustworthy
world model ($\mathbf{pre}_{1.2}$) because we have achieved $W\models\phi$
\cite{Gleirscher2021-VerifiedSynthesisSafety,%
  Gleirscher2020-SafetyControllerSynthesis} by probabilistic model
checking.
Note that errors in $\mathcal{E}$ and $\mathcal{C}$ can be iteratively identified
by validating $\phi$ (e.g., completeness and vacuity
checks) and re-checking $W\models\phi$.

$\mathbf{post}_{1}$: Here, our goal is to preserve trustworthiness $\phi$ in
the supervisor behavior $\mathsf{C}\in\mathcal{C}$ selected according to
\Cref{sec:deriv-test-refer}, namely,
$(\mathsf{E};\mathsf{C})^{\star}\models\phi'$ established by the policy
synthesis facility of the model checker ($\mathbf{post}_{1.2}$).  $\phi'$ with
$\phi\Rightarrow\phi'$ contains only those properties that can be
checked of DTMCs ($\mathbf{post}_{1.1}$).  Recall that $\mathsf{E}$ results from
converting indeterminacy in $\mathcal{E}$ into probabilistic choice, thus
qualitatively preserving all behavior of $\mathcal{E}$ in $\mathsf{E}$ while
$\mathsf{C}$ results from fixing all non-deterministic choices in
$\mathcal{C}$.
$\mathsf{C}$ will thus be deterministic and, because of being exposed to
all behaviors of $\mathsf{E}$, be able to deal with all
environments producible by $\mathcal{W}$ from $s_0$ (or
$\overline{s}$).  Given $P_1$, we can now see, that the weakest
precondition of $\mathbf{post}_{1}$ under $P_1$ is implied by $\mathbf{pre}_{1}$,
formally, 
\begin{equation}
  \label{eq:ht-synthesis-discharged}
  \{\phi\; \text{is
  complete}, W\models\phi\}
\Rightarrow \mathrm{\mathbf{\tilde{w}p}}(P_1,\{ \phi'\;\text{is complete},
  (\mathsf{E};\mathsf{C})^{\star}\models\phi'\})\;.
\end{equation}
By Proposition~\eqref{eq:ht-synthesis-discharged}, we have now
established Proposition~\eqref{eq:ht-synthesis}.

\subsection{Assurance of $P_2$: Reference Generation}
\label{sec:arg-refgen}

For $P_2$, we need to establish
\begin{equation}
  \label{eq:ht-refgen}
  \{\mathbf{pre}_{2}\} P_2 \{\mathbf{post}_{2}\}\;.
  \tag{G2}
\end{equation}

$\mathbf{pre}_{2}$: We require that
\begin{inparaenum}[(i)]
\item  $\mathsf{C}$ can be \emph{trusted}, which is 
  implied by $\mathbf{post}_{1.2}$.
\item $\mathsf{C}$ is expected to \emph{handle combinations
    of critical events}, which is established by $P_1$ generating
  commands for each factor in $F$.
\item $\mathsf{C}$ has to be deterministic, which is guaranteed by policy
  synthesis in $P_1$.
\item Supervisor
  behavior $\mathsf{C}$ is a function of $F$ and $I$,
  formally, $I\subset V$ is such that
  $\forall s,s'\in\mathcal{S}\colon (\forall v\in F\cup I\colon
  s(v)=s'(v)) \Rightarrow \mathsf{C}(s)=\mathsf{C}(s')$.
\end{inparaenum}
Note that we have now established
$\mathbf{post}_{1}\land\mbox{(iv)}\Rightarrow\mathbf{pre}_{2}$.

$\mathbf{post}_{2}$:
\begin{inparaenum}[(i)]
\item The equivalence
  $(\mathsf{E};\mathsf{C})^{\star}=(\mathsf{E};\mathcal{R})^{\star}$
  ($\mathbf{post}_{2.1}$) follows from the 1-to-1 translation of DTMC to SFSM
  transitions in~$P_2$ (\Cref{sec:test-ref-gen}).
\item Factor handling completeness ($\mathbf{post}_{2.2}$) and determinism
  ($\mathbf{post}_{2.3}$) of $\mathcal{R}$ are also maintained by this
  translation and $\mathbf{post}_{1}$.  
\item Congruence of $s_0$ and $\overline{s}$ ($\mathbf{post}_{2.4}$)
  follows from the definition of $\mathcal{R}$ (\Cref{sec:test-ref-gen}),
  from $s_0$ being the initial state of $W$ and
  $(\mathsf{E};\mathsf{C})^{\star}$, and from the fact that $\mathsf{E}$ cannot
  change~$F$.
\item \textsc{Yap}\xspace produces $\mathcal{R}$ in the \texttt{libfsmtest}\xspace input format
  ($\mathbf{post}_{2.5}$).
\end{inparaenum}
Note that $\mathcal{R}$ can be trusted
($(\mathsf{E},\mathcal{R})^{\star}\models\phi'$) but is not yet
input-complete as self-loop transitions (i.e., \emph{idle} actions)
are both ignored and not introduced by $P_2$; this completion is done
in $P_4$.

We can now derive the weakest precondition of $\mathbf{post}_{2}$ under $P_2$ as
follows:
\begin{align}
  \label{wp:refgen}
  \mathrm{\mathbf{\tilde{w}p}}(P_2,\{\mathbf{post}_{2.1}, \dots, \mathbf{post}_{2.5}\}) =
  \{\;
  \mbox{1-to-1 translation from}\;\mathsf{C}\;\mbox{to}\;\mathcal{R}
  &\;,
  \\
  \mathsf{C}\mbox{'s transitions implement the risk factors in}\;F
  &\;,
  \\
  \mathsf{C}\;\mbox{is also deterministic over}\;
  \mathcal{S}|_{F\cup I}
  &\;,
  \\
  s_0\in\mathcal{S}
  &\;\}\;.
\end{align}
Because conjuncts (i-iv) of $\mathbf{pre}_{2}$ imply the derived
precondition parts, we have 
\begin{equation}
  \label{eq:ht-refgen-discharged}
  \mathbf{post}_{1}\land\mbox{(iv)} \Rightarrow \mathbf{pre}_{2} \Rightarrow \mathrm{\mathbf{\tilde{w}p}}(P_2,
    \{\mathbf{post}_{2.1}, \dots, \mathbf{post}_{2.5}\})\;.
\end{equation}
By obtaining Proposition~\eqref{eq:ht-refgen-discharged}, we have
established Proposition~\eqref{eq:ht-refgen}.

\subsection{Assurance of $P_3$: Code Generation}
\label{sec:arg-codegen}

For $P_3$, we need to establish
\begin{equation}
  \label{eq:ht-codegen}
  \{\mathbf{pre}_{3}\} P_3 \{\mathbf{post}_{3}\}\;.
  \tag{G3}
\end{equation}
$\mathbf{pre}_{3}$ includes $\mathbf{post}_{1}$, established by $P_1$ according to
Proposition~\eqref{eq:ht-synthesis-discharged}, and the definition of
the sets $F$ (implied by $\mathbf{pre}_{1}$), $I$, and $O$
(implied by $\mathbf{pre}_{2}$).

$\mathbf{post}_{3}$: We require from $P_3$ to obtain an implementation
$C$ that can be integrated with the test harness ($\mathbf{post}_{3.1}$)
and is statically analyzable ($\mathbf{post}_{3.2}$), allowing simple checks in
$P_4$ using a custom lex/yacc parser or Unix text processing tools.
For $\mathbf{post}_{3.1}$, \textsc{Yap}\xspace generates the I/O wrapper based on a
\texttt{libfsmtest}\xspace template and the variable sets $F$, $I$, and
$O$.  For $\mathbf{post}_{3.2}$, \textsc{Yap}\xspace produces C++ code adhering to the
structure in \Cref{alg:codepattern}, assuring that no pre-processing
directives are used and the branching structure is flat and simple,
see \Cref{lst:ctrimpl}.
Trivially, $\mathbf{post}_{1}\Rightarrow\mathbf{pre}_{3}$ and the weakest
precondition of $\mathbf{post}_{3}$ under $P_3$ is:
\begin{align}
  \label{wp:codegen}
  \mathrm{\mathbf{\tilde{w}p}}(P_3,\{\mathbf{post}_{3.1}, \mathbf{post}_{3.2}\}) =
  \{\;
  (F,I,O)\;\mbox{defined}
          \;
  \}\;. %
\end{align}
Because the conjuncts of $\mathbf{pre}_{3}$ imply the derived precondition, we have 
\begin{equation}
  \label{eq:ht-codegen-discharged}
  \mathbf{post}_{1} \land \mathbf{pre}_{2} \Rightarrow \mathbf{pre}_{3} \Rightarrow \mathrm{\mathbf{\tilde{w}p}}(P_3,\mathbf{post}_{3})\;.
\end{equation}
Having verified Proposition~\eqref{eq:ht-codegen-discharged}, we have
established Proposition~\eqref{eq:ht-codegen}.

\subsection{Assurance of $P_4$: Test Suite Generation}
\label{sec:arg-testderiv}

For $P_4$, we need to establish
\begin{equation}
  \label{eq:ht-ts}
  \{\mathbf{pre}_{4}\} P_4 \{\mathbf{post}_{4}\}\;.
  \tag{G4}
\end{equation}

$\mathbf{pre}_{4}$: As specified in \Cref{tab:workflow-ag}, the precondition to be established for
task $P_4$ consists of several main conjuncts. 

$\mathbf{pre}_{4.1}$ requires the applied testing theory to be correct, when
applied with input equivalence classes specified by guard conditions
and with the H-Method for generating the intermediate FSM test suite
$\mathrm{TS}^{\mathit{fsm}}$ from $\mathfrak{A}(\mathcal{R})$
(\Cref{sec:test-strategy}). The correctness of the input equivalence
class construction has been proven by Huang et
al.~\cite{peleskasttt2014}.  They have also shown that \emph{any}
complete FSM test generation method can be applied for calculating
$\mathrm{TS}^{\mathit{fsm}}$, and the resulting SFSM test suite will always be
complete as well~\cite{Huang2017}.  Finally, Dorofeeva et
al.~\cite{DBLP:conf/forte/DorofeevaEY05} have proven the completeness
of the H-Method. These facts ensure the validity of $\mathbf{pre}_{4.1}$.

$\mathbf{pre}_{4.2}$ requires that the prerequisites for applying this test
theory are fulfilled by the reference model $\mathcal{R}$ and the
implementation $C$. The four prerequisites to be established
are defined and ensured as follows.  (i) It has to be shown that the
reference model $\mathcal{R}$ can be interpreted as a deterministic
reactive I/O state transition system ($\mathbf{pre}_{4.2.1}$); this is a
variant of Kripke structures that is used as the semantic basis for
models in Huang et al.~\cite{peleskasttt2014}.  SFSMs represent
special cases of reactive I/O state transition systems.  The
determinism of $\mathcal{R}$ is ensured by $\mathbf{post}_{2}$.  (ii) It has to be
justified that the true behavior of $C$ can be interpreted as
a deterministic reactive I/O state transition system
($\mathbf{pre}_{4.2.2}$). This is easy to see, since the code generator creates
$C$ with a structure where a main loop evaluates which guard
condition can be applied and executes the corresponding transition
changing the internal control state and setting finite-valued
outputs. By static code analysis, it is shown that $C$ uses
exactly the same guard conditions as $\mathcal{R}$. This can be performed
with Unix core utilities like {\tt grep}, {\tt sed}, and {\tt
  sort}\footnote{\url{https://man7.org/linux/man-pages/dir_section_1.html}}. Thus,
$C$ is deterministic since $\mathcal{R}$ is deterministic.  (iii)
Next, it has to be shown that the implementation $C$ does not
possess more than $m$ control states ($m\ge n$, where
$n\leq|S|$ after minimization of $\mathcal{R}$)
($\mathbf{pre}_{4.2.3}$).  The number of states in
  $\mathfrak{A}(\mathcal{R})$ can be directly determined from a description
file containing all such states. Again, a
simple static analysis shows that the code uses the same
control states, so $m = n$.  (iv) Finally, it has to be shown that
each guard condition represents a single input equivalence class
($\mathbf{pre}_{4.2.4}$). Indeed, this follows from the fact that all
guards are mutually exclusive, which has already been established by
$\mathbf{post}_{2}$.

$\mathbf{pre}_{4.3}$: Instead of verifying the H-Method algorithm in
$\texttt{libfsmtest}\xspace$, we check the generated test suite $\mathrm{TS}^{\mathit{fsm}}$ for
consistency with the test suite specification of the H-Method. If the
check fails, the algorithm can be fixed, and the suite can be created
again. The check is automatically performed by a \emph{test suite
  validator} $\text{Val}_H$.  As explained in \Cref{sec:tq},
this approach requires to qualify $\text{Val}_H$.  Qualifying
$\text{Val}_H$ instead of the H-Method algorithm has the
advantage that the former has a significantly simpler implementation
than the latter. Also, any future optimizations of the test algorithm
will not affect the \emph{tool qualification (TQ)} of $\text{Val}_H$,
because the checks performed on the generated test suites do not
depend on the library, but only on the theory.
Summarizing, 
\begin{align}
  \mathbf{pre}_{4} \equiv \mathbf{pre}_{4.1} \wedge \mathbf{pre}_{4.2} \wedge \mathbf{pre}_{4.3}, & \tag{Definition}  
  \\
  \mathbf{pre}_{4.3} \equiv \mathbf{pre}_{4.2.1} \wedge \mathbf{pre}_{4.2.2} \wedge \mathbf{pre}_{4.2.3} \wedge \mathbf{pre}_{4.2.4},
  & \tag{Definition}  
\end{align}
and the validity of $\mathbf{pre}_{4.1}$, $\mathbf{pre}_{4.2}$, $\mathbf{pre}_{4.2.1}$, $\mathbf{pre}_{4.2.2}$, $\mathbf{pre}_{4.2.3}$, and
$\mathbf{pre}_{4.2.4}$ has been ensured. The validity of $\mathbf{pre}_{4.3}$ is explained below.
Therefore, the validity of precondition $\mathbf{pre}_{4}$ is ensured as well.

$\mathbf{post}_{4}$: The post-condition to be established   is that the   test suite
$\mathrm{TS}^{\mathit{fsm}}$ generated by $P_4$ according to the steps described in \Cref{sec:test-strategy} 
is complete for testing  any FSM over alphabets $\mathfrak{A}({\mathcal{R}})_{I}, \mathfrak{A}({\mathcal{R}})_{O}$ 
with at most $m$ states  
against $\mathfrak{A}({\mathcal{R}})$, provided that $\mathbf{pre}_{4}$ holds and $\text{Val}_H$ does not
indicate a generation error.

The weakest precondition ensuring $\mathbf{post}_{4}$ under $P_4$ is
\begin{align}
  \label{wp:ts}
  \mathrm{\mathbf{\tilde{w}p}}(P_4,\{\mathbf{post}_{4}\}) =
  \{\;
  \text{Testing theory is correct}    &\;,  
  \\
  \text{Prerequisites for applying testing theory are fulfilled}    &\;,  
  \\
  \text{$\mathrm{TS}^{\mathit{fsm}}$ is complete if $\text{Val}_H$  indicates no error}&\;\}    
\end{align}
Obviously, 
\begin{equation}
  \label{eq:ht-ts-discharged}
  \mathbf{pre}_{4} \Rightarrow \mathrm{\mathbf{\tilde{w}p}}(P_4,\mathbf{post}_{4})\;.
\end{equation}
By Proposition~\eqref{eq:ht-ts-discharged}, we have established
Proposition~\eqref{eq:ht-ts}.

\subsection{Assurance of $P_5$: Conformance Testing}
\label{sec:arg-conftest}

For $P_5$, we need to establish
\begin{equation}
  \label{eq:ht-test}
  \{\mathbf{pre}_{5}\} P_5 \{\mathbf{post}_{5}\}\;.
  \tag{G5}
\end{equation}

$\mathbf{pre}_{5}$:
The post-condition to be established (see \Cref{tab:workflow-ag}) states that 
$C$ passes the test suite $\mathrm{TS}$ 
if and only if it is observationally equivalent to
$\mathcal{R}$. As explained in \Cref{sec:conftesting}, 
the test wrapper with embedded supervisor (denoted by $\texttt{TW}[C]$) 
implements an FSM over the same input and output alphabet 
as $\mathfrak{A}({\mathcal{R}})$. The test executable  $\texttt{TH}[\texttt{TW}[C]]$
runs test suite $\mathrm{TS}^{\mathit{fsm}}$ against $\texttt{TW}[C]$; this is equivalent to running
$\mathrm{TS}$ against $C$, provided that the wrapper implements the alphabet mappings $\gamma,\omega$ 
correctly. 
Therefore, we require that $\mathrm{TS}^{\mathit{fsm}}$ is complete ($\mathbf{pre}_{5.1}$), which has already been established by 
$\mathbf{post}_{4}$. The additional conjuncts of $\mathbf{pre}_{5}$ are tool-related: 
\begin{inparaenum}[(i)]
\item The test wrapper $\texttt{TW}$ is correct---this is ensured by two sub-conditions.
\begin{inparaenum}
\item[$\mathbf{pre}_{5.2}$:] $\texttt{TW}$ correctly implements $\gamma$ as defined in \Cref{sec:conftesting}.
\item[$\mathbf{pre}_{5.3}$:] $\texttt{TW}$ correctly implements $\omega$ as defined in \Cref{sec:conftesting}.
\end{inparaenum}
\item The test harness $\texttt{TH}$ is correct---this is ensured by three sub-conditions.
\begin{inparaenum}
\item[$\mathbf{pre}_{5.4}$:] $\texttt{TH}$  skips no test cases from $\mathrm{TS}^{\mathit{fsm}}$.
\item[$\mathbf{pre}_{5.5}$:] $\texttt{TH}$  neither skips nor adds nor changes
  inputs in a test case $t\in\mathrm{TS}^{\mathit{fsm}}$.
\item[$\mathbf{pre}_{5.6}$:] $\texttt{TH}$ assigns verdict PASS to execution
  of $t$ if and only if the I/O trace produced by
  $\texttt{TW}[C]$ for $t$ is in the language
  $L(\mathfrak{A}({\mathcal{R}}))$ of $\mathfrak{A}({\mathcal{R}})$.
\end{inparaenum}
\end{inparaenum}

Preconditions $\mathbf{pre}_{5.2}$ and $\mathbf{pre}_{5.3}$ are ensured by comprehensive
TQ tests (as explained below in \Cref{sec:tq}) qualifying the test
wrapper. Precondition $\mathbf{pre}_{5.6}$ is ensured analogously for the
checker component of the test harness $\texttt{TH}$.  Preconditions
$\mathbf{pre}_{5.4}$ and $\mathbf{pre}_{5.5}$ are ensured by artifact-based TQ
(\Cref{sec:tq}): the test cases actually executed and their associated
test steps are documented in a test execution log that is compared to
the test suite $\mathrm{TS}^{\mathit{fsm}}$.  The comparison is performed by an
\emph{execution log validator} $\text{Val}_{log}$.  The logging
component of $\texttt{TH}$ and the log validator $\text{Val}_{log}$
are qualified by means of comprehensive TQ tests.

$\mathbf{post}_{5}$: 
The test theory explained in \Cref{sec:arg-testderiv} implies that 
$\mathbf{post}_{5}$ is equivalent to the following alternative post-condition.
\begin{quote}
$\mathbf{post}_{5}'$: $\texttt{TW}[C]$ passes the test suite $\mathrm{TS}^{\mathit{fsm}}$ 
if and only if it is observationally equivalent to $\mathfrak{A}({\mathcal{R}})$.
\end{quote}
Of course,  $\mathbf{post}_{5}$ and $\mathbf{post}_{5}'$ are only equivalent if the
wrapper correctly implements the
alphabet mappings $\gamma$ and $\omega$ (\Cref{sec:conftesting}). 
We know from $\mathbf{post}_{4}$ that $\mathrm{TS}^{\mathit{fsm}}$ is complete, so it is a good candidate for
checking observational equivalence. Additionally, however, it needs to be ensured that the
test harness $\texttt{TH}$ executes the test suite correctly and performs the correct
checks of observed $C$-reactions against  reactions 
expected according to $\mathfrak{A}({\mathcal{R}})$.
Consequently, the weakest precondition of $\mathbf{post}_{5}$ under $P_5$ is
\begin{align}
  \label{wp:test}
  \mathrm{\mathbf{\tilde{w}p}}(P_5,\{\mathbf{post}_{5}'\}) =
  \{\;
  \text{$\mathrm{TS}^{\mathit{fsm}}$ is complete}  &\;,  
  \\
  \text{$\texttt{TW}$ implements $\gamma$ correctly}  &\;,  
  \\
  \text{$\texttt{TW}$ implements $\omega$ correctly}  &\;,  
  \\
  \text{$\texttt{TH}$ skips no test cases of $\mathrm{TS}^{\mathit{fsm}}$}  &\;,  
  \\
  \text{$\texttt{TH}$ neither skips nor adds nor changes inputs in a test case}  &\;,  
  \\
 \text{Test case $t$ passes iff $\texttt{TW}[C](t)\in L(\mathfrak{A}({\mathcal{R}}))$}  &\;
  \}  
\end{align}
Because the conjuncts of $\mathbf{pre}_{5}$ are equivalent to these preconditions, we have 
\begin{equation}
  \label{eq:ht-test-discharged}
  \mathbf{pre}_{5.1}\land\dots\land \mathbf{pre}_{5.6}  \Rightarrow \mathrm{\mathbf{\tilde{w}p}}(P_5,\mathbf{post}_{5})
\end{equation}
By Proposition~\eqref{eq:ht-test-discharged}, we have established
Proposition~\eqref{eq:ht-test}.

\subsection{Tool Qualification} 
\label{sec:tq}

As described above, the soundness of the workflow establishing the
refinement chain \eqref{eq:ref-chain} depends on tools automating
critical verification steps.  Following the applicable standards for
safety-critical systems development, this workflow requires tool
  qualification.  For TQ-related considerations, we apply the
avionic standard RTCA DO-178C with annex DO-330~\cite{DO178C,DO330},
because this is currently the most specific and strict standard, as
far as TQ is concerned.  Fulfilling the TQ requirements specified
there implies compatibility with the requirements for support tools
according to IEC~61508~\cite{iec61508} and ISO~10218~\cite{ISO10218}.

Standards for the use of automation tools in development and
verification (e.g., RTCA DO-178C) offer three options to ensure that
the tool-produced artifacts (e.g.,~object code, test suites, test
execution results) are correct.  (1) If an artifact is not verified by
any other means, the tool needs to be qualified. (2) If an artifact is
verified manually by a systematic review or inspection, no TQ is
required. (3) If an artifact is verified by an automated checker
replacing the manual review/inspection procedure, then the checker
needs to be qualified~\cite{peleska2012c}.  We call this
\emph{artifact-based TQ}, since the tool is ``re-qualified'' every
time it produces a new artifact (i.e.,~a new test suite in our case).

For either test automation tools or associated artifact checkers, the so-called \emph{tool qualification level} TQL-4 specified in \cite[Section~12.2.2]{DO178C} applies:
this level is intended for tools that may not produce errors in software to be deployed in the target system, but whose failures may prevent the detection of errors in target software.
TQL-4 requires a documented development life cycle for the tool, a
comprehensive requirements specification, and a verification that
these requirements are fulfilled. Verification can be performed by
reviews, analyses (including formal verification), and tests. Formal
verification of the tool alone cannot replace SW/HW integration tests
of the tool on the platform where it is deployed. 
For TQL-4 tools to be applied for the verification of target software of highest criticality (as discussed in this paper), the TQ tests need not only cover all requirements, but also the code with 100\% MC/DC  coverage~\cite[p.~114]{DO178C}.

We are aware that the workflow stages $P_1$ and $P_2$ are subjected to
TQ as well.  However, the qualification of model checkers and their
results~\cite{Wagner2017-QualificationModelChecker} in $P_1$ is more
complicated, part of ongoing research, and thus out of scope here.

\vspace{-.5em}

\vspace{-.5em}

\section{Related Work and Discussion}
\label{sec:related-work}
\label{sec:discussion}

The approaches in \cite{Orlandini2013-ControllerSynthesisSafety,%
  Bersani2020-PuRSUEspecificationrobotic} from collaborative robotics
are perhaps closest to our workflow ($P_1$) as they include platform
deployment ($P_3$).  While they focus on the synthesis of overall
robot controllers, we focus on supervisors but with a testing stage
reassuring implementation correctness.
Villani et al.~\cite{Villani2019-Integratingmodelchecking} integrate
quantitative model checking~(with
\textsc{Uppaal}~\cite{Behrmann2004-TutorialUppaal}) with conformance
testing and fault injection.  The authors advocate cross-validation of
\textsc{Uppaal} and FSM models.  Our approach differs from theirs in
two ways: (i) SFSMs do not require cross-validation, since they are
generated from a world model validated by model checking.  (ii) We do
not need fault injection for testing, since our complete test strategy
corresponds to a formal code verification by model checking.

Research in complete testing methods is a very active
field~\cite{Petrenko:2012:MTS:2347096.2347101}.  We applied the
H-Method~\cite{DBLP:conf/forte/DorofeevaEY05} in $P_4$ because (i) it
produces far fewer test cases than the classical
W-Method~\cite{chow:wmethod}, but (ii) it allows for an intuitive test
case selection, facilitating the qualification of the test case
generator (\Cref{sec:test-strategy}).  If the objective of a test
campaign is to provide complete suites with a minimal number of test
cases then the SPYH-Method~\cite{DBLP:conf/icst/SouchaB18} should be
preferred to the H-Method.

Hazard/failure-oriented %
testing~\cite{Gleirscher2011-HazardbasedSelection%
  ,Lesage2021-SASSISafetyAnalysis} and requirements falsification
based on negative
scenarios~\cite{Uchitel2002-Negativescenariosimplied%
  ,Gleirscher2014-BehavioralSafetyTechnical%
  ,Stenkova2019-GenericNegativeScenarios} are useful if $\mathcal{R}$ is
not available or needs to be validated and revised.  In contrast, our
approach is \emph{complete} once $\mathcal{R}$ is validated, that is, any
deviation from $\mathcal{R}$ detectable by such techniques is also
uncovered by at least one test case generated by our approach.
Moreover, our approach is usable to test supervisor robustness without
a realistic simulator for~$W$.

The soundness argument for our workflow  
(\Cref{sec:workflow-qualif-concept}) relies on proofs of completeness
of the test suites generated with the testing theories for showing
$\mathcal{R}=C$.  Such proofs can be mechanized using proof
assistants~\cite{DBLP:conf/pts/SachtlebenHH019}.  It is also possible
to generate test generation algorithms from the proof tool and prove
their correctness as a minor extension of the testing
theory~\cite{DBLP:conf/pts/Sachtleben20}.  This could simplify the
tool qualification argument in \Cref{sec:arg-testderiv}.  However,
some kind of TQ  argument will still be necessary, 
 because proven algorithms do not guarantee
correctness of their execution on a target platform (e.g., a PC or
cloud server), where additional errors might be produced due to
inadequate address or integer register sizes.

\vspace{-.5em}

\section{Conclusions}
\label{sec:conclusion}

\vspace{-.5em}

We proposed a rigorous workflow for the sound development (i.e., the
\emph{verified synthesis and code generation}) of supervisory
discrete-event controllers enforcing safety properties in human-robot
collaboration and other autonomous system applications.
The novelty of this workflow consists in
\begin{inparaenum}[(i)]
\item the generation of a test reference model whose completeness and
  correctness is established by a refinement relation to a validated
  world model,
\item the application of complete model-based testing methods in
  combination with static analysis to obtain a conformance
  \emph{proof} of the safety supervisor code, and
\item an explanation of how the tools involved and the artifacts they
  produce can be qualified according the most stringent requirements
  from standards for safety-critical systems development.
\end{inparaenum}
We employ Hoare logic and weakest precondition calculus (at a
meta-level rather than at the artifact level) to establish soundness
of our workflow and use goal structuring notation to structure and
visualize the complex verification and validation argument to obtain
certification credit.  The workflow is supported by a tool chain:
\textsc{Yap}\xspace~\cite{Gleirscher2020-YAPToolSupport} and a stochastic
model checker~(e.g., \textsc{Prism}
\cite{Kwiatkowska2011-PRISM4Verification}) for Markov decision process
generation and verification, \textsc{Yap}\xspace for test reference and code
generation, and \texttt{libfsmtest}\xspace~\cite{libfsmtest} for test suite
derivation and execution.

To test the integrated HW/SW-system (i.e., robot, welding machine,
safety supervisor and simulation of human interactions), we will embed
our approach into a more general methodology for verification and
validation of autonomous systems, starting at the module level
considered here, and ending at the level of the integrated overall
system~\cite{DBLP:journals/corr/abs-2110-12586}.

\bibliographystyle{splncs04}
\bibliography{} %

\end{document}